\documentclass[english,twocolumns,prb]{revtex4}
\usepackage{babel}
\usepackage{amssymb,amsmath,amsfonts}
\usepackage{graphicx}
\usepackage{natbib}
\usepackage[usenames,dvipsnames]{color}

\usepackage[mathlines]{lineno}

\def\ket#1{|\mbox{$#1$}\rangle}

\begin{document}

\title{Hybrid light-matter networks of Majorana zero modes}

\author{L. C. Contamin$^{1}$, M.R. Delbecq$^{1}$, B. Dou\c cot$^{2}$, A. Cottet$^{1}$ and T. Kontos$^{1}$}
\affiliation{Laboratoire de Physique de l'Ecole normale sup\'{e}rieure, ENS, Universit\'{e} PSL, CNRS, Sorbonne Universit\'{e}, Universit\'{e} Paris-Diderot, Sorbonne Paris Cit\'{e}, Paris, France, Laboratoire de Physique Th\'eorique et des Hautes Energies, CNRS, Sorbonne Paris Cit\'{e}, France }

%\linenumbers
\begin{abstract}
Topological excitations, such as Majorana zero modes, are a promising route for encoding quantum information. Topologically protected gates of Majorana qubits, based on their braiding, will require some form of network. Here, we propose to build such a network by entangling Majorana matter with light in a microwave cavity QED setup. Our scheme exploits a light-induced interaction which is universal to all the Majorana nanoscale circuit platforms. This effect stems from a parametric drive of the light-matter coupling in a one-dimensional chain of physical Majorana modes. Our setup enables all the basic operations needed in a Majorana quantum computing platform such as fusing, braiding, the crucial T-gate, the read-out and, importantly, the stabilization or correction of the physical Majorana modes.
\end{abstract}

\maketitle

Majorana quasiparticles in condensed matter systems have been the subject of intense experimental work for almost a decade\cite{Albrecht:16,Gul:18,Desjardins:19,Ren:19,Fornieri:19}, for their potential in defining topologically protected qubits and gates\cite{Alicea2011}. However, experimental realizations have not succeeded so far in measuring the expected non-abelian statistics of these exotic excitations. Several protocols have been proposed for performing such advanced experiments through electronic transport measurements\cite{Alicea2011,Hassler2011, VanHeck2012, Hyart2013, You2014, Aasen2015a, Vijay2016, Karzig2017, Knapp2019}. They all require a microscopic control and fine tuning of the experimental platforms, a 2D or at least a network geometry and an invasive transport based read-out.

Cavity photons have appeared as a major toggle for manipulating, coupling and reading out the quantum state of superconducting circuits\cite{Clerk2010,Hays2020}. However, the direct application of Circuit QED techniques to Majorana fermions is hindered by their self-adjoint property which forbids a direct energy exchange between an isolated Majorana doublet and a cavity\cite{Dartiailh2017}. It was proposed to probe the presence and parity of a given Majorana doublet by observing charge transitions to supplementary states\cite{Cottet2013, Dartiailh2017,Dmytruk2015} or by using a charge sensitive Josephson circuit\cite{Muller2013}. In principle, one can also detect the dynamical phase resulting from the braiding of Majorana fermions by probing the cavity field\cite{Trif2019b}. However, the above proposals are elusive regarding the manipulation and coupling of Majorana states through the photonic degree of freedom. This is why circuit QED could not be envisionned as a full platform for performing all the requested operations for fusing and braiding the MBSs, so far.

In this paper, we propose a hybrid Majorana-cavity platform which fills these gaps. We show that by modulating the Majorana-photon coupling at the cavity frequency, one can fuse and braid two MBSs or perform T-gates in a four to six MBSs \textit{linear} chain. This resource is obtained because the modulation produces an effective 2D network out of a 1D chain.  Finally, we show how we can preserve the topological protection of the Majorana modes using an active stabilization based on the joint action of the cavity photons and the modulation of the coupling.

\begin{figure}
    \centering
    {\small \centering\includegraphics[width=0.5\linewidth,angle=0]{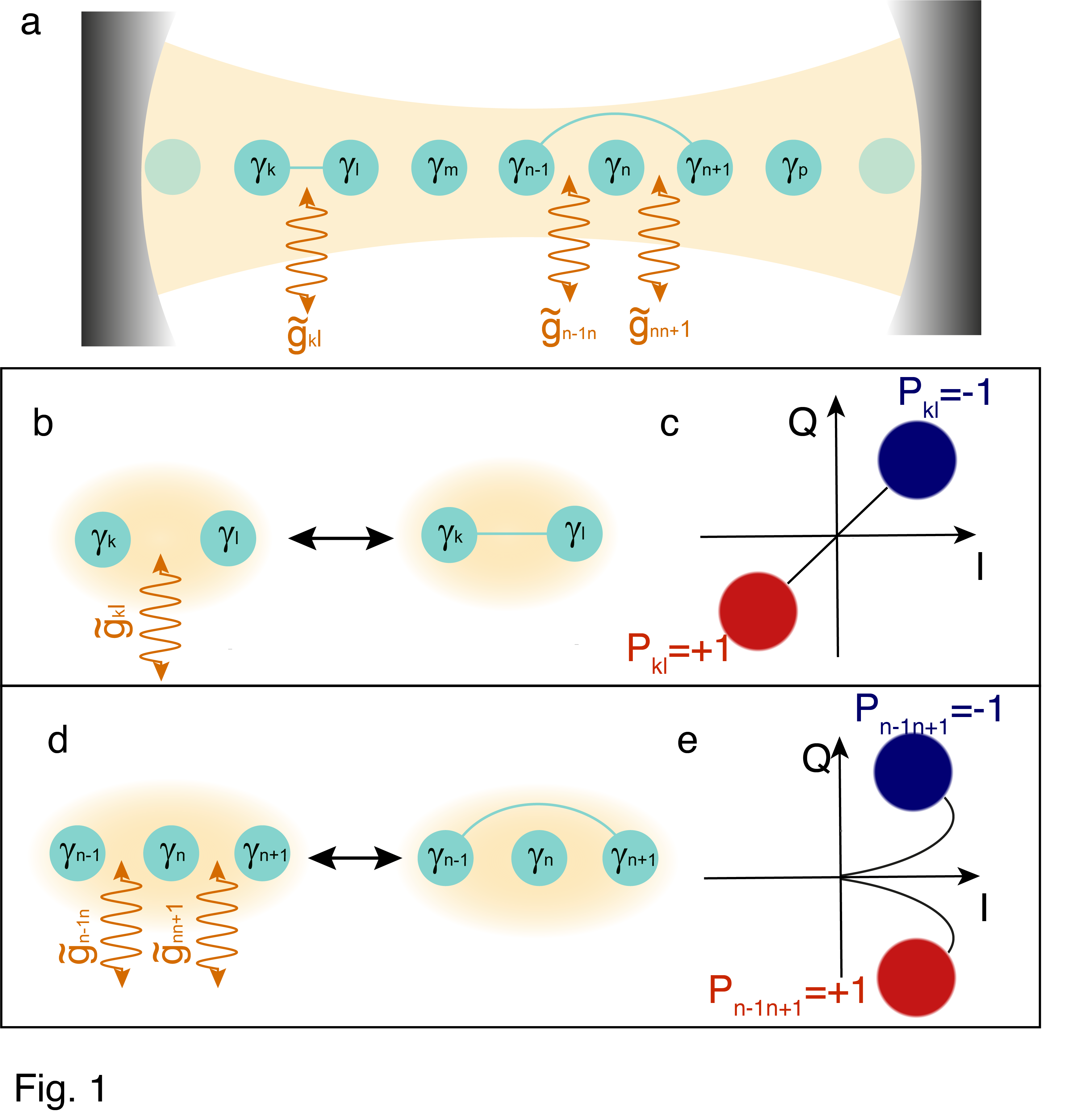}}
    \caption{\textbf{Hybrid light-matter network of Majorana zero modes.} \textbf{a} We consider a chain of  Majorana quasiparticles $\hat{\gamma}_{j}$ embedded inside a microwave cavity. The cavity is represented the two mirrors (shaded black) concentrating a photonic field (light yellow) around the circuit. The Majorana modes are represented as turquoise balls. The microwave drive of each section $(j,j+1)$ is represented as orange vertical wavy lines. The resulting effective interaction is represented as turquoise lines, turning the chain into an elementary network suitable for fusing, braiding and stabilizing Majorana modes. \textbf{b} Nearest neighbour interaction and its signature in the trajectory of a coherent state in the quadratures I-Q plane of the cavity field. \textbf{c} Second nearest neighbour interaction and its signature in the trajectory of a coherent state in the quadratures I-Q plane of the cavity field.}
    \label{fig:figure1}
\end{figure}

We consider a linear chain of MBSs hosted in a nanoconductor, as represented in figure 1a. The nanoconductor may be implemented in various physical platforms. It is capacitively coupled to a microwave cavity, which we describe as a single photonic mode $\hat{a}$ with frequency $\omega_{c}/2\pi$. Each MBS is associated with a self-adjoint creation operator $\hat{\gamma}_{i}$ as depicted in figure 1a. A small overlap between two neighboring MBSs gives rise to energy splittings $\epsilon_{jj+1}$, which are exponentially suppressed with the distance between the two MBSs. The low-energy effective Hamiltonian of the system can be written as\cite{Cottet2017}:
\begin{equation}\label{eq:ham}
 \bar{H}_{chain}=\sum_{j}  g_{jj+1}(\hat{a}+\hat{a}^{\dagger})i\hat{\gamma}_{j}\hat{\gamma}_{j+1}+\sum_{j} \epsilon_{jj+1}i\hat{\gamma}_{j}\hat{\gamma}_{j+1}+\hbar\omega_{c}\hat{a}^{\dagger}\hat{a}
\end{equation}
One can associate to each Majorana pair $(j,k)$ a topological charge with a parity operator $\hat{P}_{jk} = i\hat{\gamma}_j\hat{\gamma}_k$. Unless otherwise specified, we assume that our chain has already well developped Majorana modes with energies $\epsilon_{jj+1}$ much smaller than $\hbar \omega_c$.

One of the main results of our work is how we can shape the above hamiltonian to manipulate, read-out and stabilize Majorana modes under a parametric drive. The electron-photon couplings can be locally modulated at microwave frequencies  $\omega_{RF}\simeq \omega_c$ through a modulation of local gate electrodes (which modulate the MBS overlap) such that: $g_{jj+1}(t) = \bar{g}_{jj+1} + \tilde{g}_{jj+1}\cos(\omega_{RF}t + \phi_{jj+1})$. In the following, we consider different types of parametric drives to implement the different Majorana operations. In all cases, we can transform the above hamiltonian into a quasi-static one by going into the rotating frame of the cavity field and/or performing a suitable dispersive unitary transformation. This gives:
\begin{equation}\label{eq:fulleffective}
    H_{eff}=\sum_{n,m}i\hat{\gamma}_{n}\hat{\gamma}_{m}f_{nm}(\hat{a}^{\dagger},\hat{a},\hat{a}^{\dagger}\hat{a})+\hbar\delta\hat{a}^{\dagger}\hat{a}
\end{equation}
where $f_{nm}$ is a linear combination of $\hat{a},\hat{a}^{\dagger},\hat{a}^{\dagger}\hat{a}$ which depends on the operation considered, and $\delta=\omega_c-\omega_{RF}$ is the detuning between the drive and the cavity. The change from 1D to 2D is one of the main resources which we exploit in this paper. As shown in figure 1b, 1c 1d and 1e, the read-out of the parity $\hat{P}_{jj+1}$, or $\hat{P}_{jj+2}$ of pairs $(j,j+1)$ or $(j,j+2)$ can be implemented by choosing appropriate gate voltage pulses (See Methods). The specific trajectories of the coherent cavity field carrying the information on the MBSs parity in the two elementary cases is shown in figure 1c and e. This information can be retrieved by measuring the field leaking out of the cavity with microwave techniques, as shown in the input-output theory section in the methods. Unless otherwise specified, we now assume that the chain considered has a given total parity. In addition, we will omit for clarity the j index of each Majorana until the discussion of the Majorana stabilization, replacing $j+1..j+6$ by $1..6$.

\begin{figure}
    \centering
    {\small \centering\includegraphics[width=0.50\linewidth,angle=0]{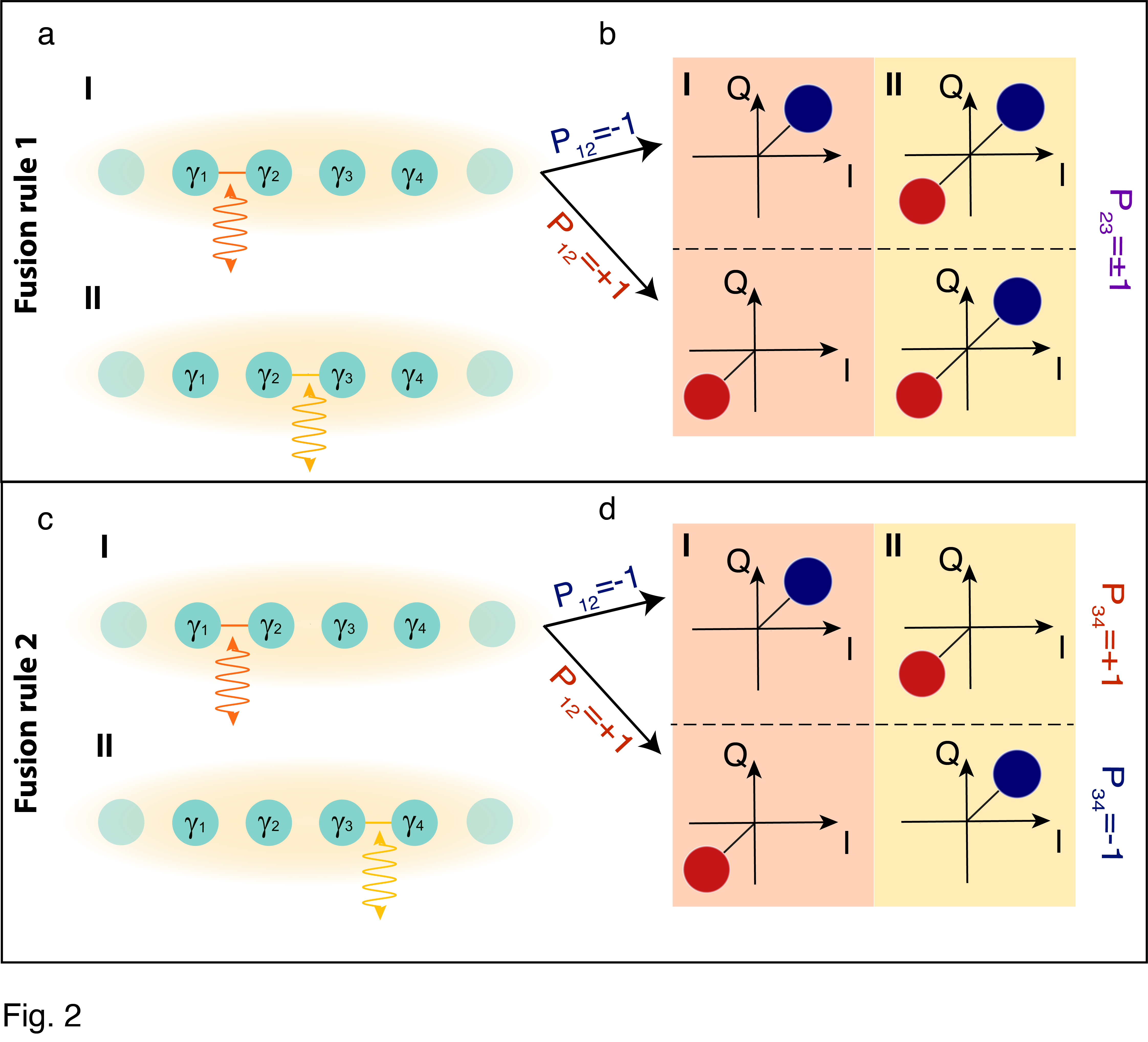}}
    \caption{\textbf{Observing the fusion rules in the field quadratures plane} \textbf{a} Pulse sequences enabling the observation of the fusion rule of two adjacent Majorana modes. The successive pulses on the two adjacent sections of the physical chain can be tracked by coherent state spots in the quadrature I-Q plane of the cavity field. \textbf{b} The fusion yields two spots of equal weight for both parities in the I-Q plane starting from either of the two parities of section $(1,2)$. \textbf{c} Pulse sequence on the two separated sections of the physical chain . \textbf{d} The fusion yields anti-correlated spots for both parities in the I-Q plane starting from either of the two parities.}
    \label{fig:figure2}
\end{figure}

The fusion of two MBSs j and k is the projective measurement of their parity $\hat{P}_{jk}$ \cite{Beenakker:19}. Measuring the coherent field spots in the quadratures I-Q plane of the cavity field is projective for separated spots like those sketched in figure 1c or 1e. Hence, our proposed setup enables to fuse pairs of MBSs. Strikingly, such a scheme also gives direct access to the fusion rules which are directly linked to the non-abelian algebra of the MBSs\cite{Beenakker:19}. The full sequence for establishing the fusion rules is represented in figure \ref{fig:figure2}2 for an odd total parity. One can measure the parity $\hat{P}_{12}$ and then $\hat{P}_{23}$ (panel a) or the parity $\hat{P}_{12}$ and then $\hat{P}_{34}$ (panel c). The results are expected to be qualitatively different whether $\hat{P}_{23}$ or $\hat{P}_{34}$ is measured. In the first case, random spots with equal weight should appear in the I-Q plane along the axis defined by the first parity measurement whereas perfectly anti-correlated spots should appear in the second case. Specifically, the second parity measurement of the sequence of figure 2a shows directly that the fusion creates an equal weight coherent superposition of states. These constitute a direct signature of the fusion rules of MBS1 and MBS2.

\begin{figure}
    \centering
    {\small \centering\includegraphics[width=0.5\linewidth,angle=0]{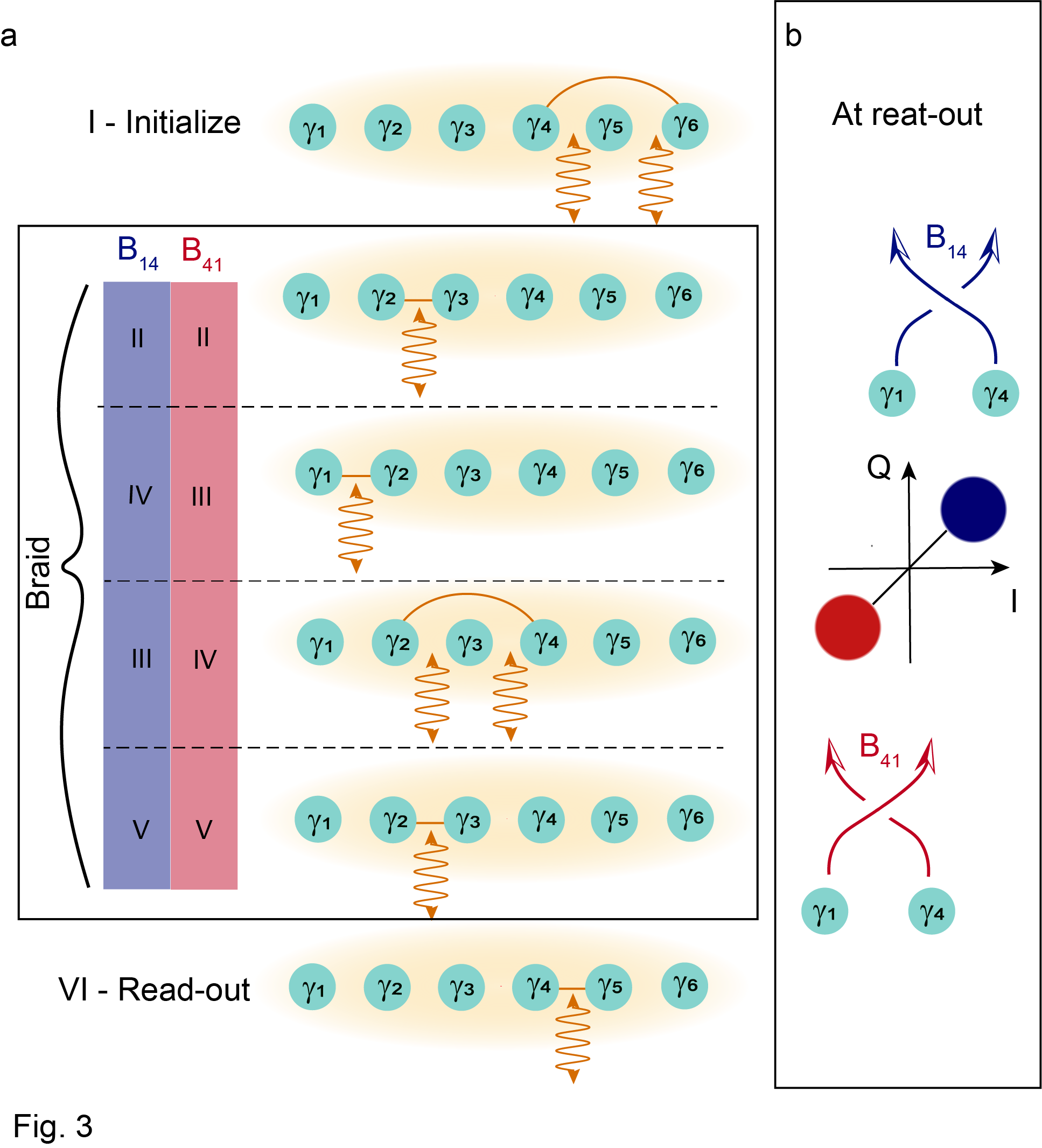}}
    \caption{\textbf{Braiding protocol in a chain of 6 MBSs} \textbf{a} Pulse sequence enabling the "clockwise" or "counterclockwise" braiding depending on the order of pulse III or IV. The first pulse is an initialization and the last pulse is the readout. \textbf{b} Result of the clockwise and counterclockwise braiding as observed in the I-Q plane trajectory of the coherent state spot (blue or red spots) after pulse VI. The qualitative difference of the cavity field in the two possible braids would be a direct observation of the non-abelian braiding.}
    \label{fig:figure4}
\end{figure}

 The braiding of two MBSs is the coherent exchange of them. Performing such an exchange in 1D is a challenge. It has been suggested  to make use of anyon teleportation by strong parity measurements\cite{Bonderson2008,Zilberberg2008,Beenakker:19} rather than moving in real space or in phase space the MBSs. These ideas have not been implemented so far. One important roadblock is that one needs to read-out the parity corresponding to distant Majorana's such as $2$ and $4$. The conventional wisdom is that this still requires a network geometry since it seems difficult to ``jump over'' the intermediate Majorana (here Majorana $3$) in a 1D setup\cite{Vijay2016}. However, this can be done thanks to the microwave cavity by using two phase shifted modulation pulses (optimally by $\pi/2$) with $\tilde{g}_{23}\neq 0$ and $\tilde{g}_{34}\neq 0$, turning effectively our 1D system into a synthetic, light-induced, 2D system (see methods eq. (8)). The braiding of MBS $1$ and $4$ can be performed by using the anyon teleportation protocol enabled by our hybrid light-Majorana platform. One has to first measure $\hat{P}_{23}$ and postselect the $+1$ eigenvalue as an initialization step, then $\hat{P}_{21}$, then $\hat{P}_{24}$ and finally $\hat{P}_{23}$ again (also postselecting the $+1$ eigenvalue, see methods section) to perform the braiding\cite{Bonderson2008,Vijay2016}. The non-abelian nature of the braiding can be directly seen by changing the order of the parity measurements $\hat{P}_{21}$ and $\hat{P}_{24}$ and obtaining different measurement outcomes for the total wave function of the chain. In the four MBSs chain, the change in the total wave function is the geometrical phase $\pm \pi/4$, which cannot be sensed directly by the cavity photons. Importantly, this geometrical phase has a measurable signature if we enlarge the chain to six MBSs, as described in figure 3, to make the clockwise and anticlockwise braiding paths interfere. For that purpose, we enrich the anyon teleportation protocol by the initialization of the state through the measurement of $\hat{P}_{46}$ and the postselection of the $\ket{0_{46}}$ parity state, starting from a two fermion-state, e.g.$\ket{1_{12}1_{34}0_{56}})$. In the latter case, this gives the initial state
$\ket{\Psi_{init}} = \frac{1}{\sqrt{2}}(\ket{1_{12}1_{34}0_{56}}+i\ket{1_{12}0_{34}1_{56}})$, which we write in the natural basis formed by the eigenstates of $\hat{P}_{12}$, $\hat{P}_{34}$ and $\hat{P}_{56}$. We restrict the discussion to the even total parity (the discussion for odd total parity is very similar). The initial state $\ket{\Psi_{init}}$ creates a superposition of two different parities in the subspace associated with the four MBSs 1 to 4. Since they live in different parity subspaces, they pick up opposite $\pi/4$ phases during the braiding operation. The choice of the initial state and the pulse sequence makes them interfere like in a polarizer/analyzer setup with birefringent media. The corresponding pulse sequences for $B_{14}$ and $B_{41}$ are displayed in figure \ref{fig:figure4}3a. After the initialization sequence, one should measure  $\hat{P}_{23}$, $\hat{P}_{21}$, then  $\hat{P}_{24}$ and then  $\hat{P}_{23}$ for $B_{41}$, postselecting the $+1$ eigenvalue. For $B_{14}$, one should measure  $\hat{P}_{23}$, $\hat{P}_{24}$, then  $\hat{P}_{21}$ and then  $\hat{P}_{23}$, postselecting the $+1$ eigenvalue.
After the braiding (see methods), we obtain the state $\ket{\Psi}_{\text{braided 14}}= \frac{e^{-i\pi/4}}{2}(\ket{0_{12}0_{34}0_{56}}+\ket{1_{12}1_{34}0_{56}}-\ket{1_{12}0_{34}1_{56}} -\ket{0_{12}1_{34}1_{56}})$ for the clockwise braiding and $\ket{\Psi}_{\text{braided 41}} =\frac{e^{i\pi/4}}{2}(\ket{0_{12}0_{34}0_{56}}+\ket{1_{12}1_{34}0_{56}}+\ket{1_{12}0_{34}1_{56}} +\ket{0_{12}1_{34}1_{56}})$ for the anti-clockwise braiding. The non-abelian character of the operation becomes therefore directly visible in the different outcomes of the coherent field spots in the I-Q plane for the parity $\hat{P}_{45}$ measurement which is carried out at the last step in our protocol (see figure 3a). The clockwise braiding corresponds to the blue spot ($\hat{P}_{45}=-1$) whereas the anti-clockwise braiding corresponds to the red spot (($\hat{P}_{45}=+1$)).

The above methods for fusion or braiding can be extended to more complex gates. In particular, the T-gate (also called $\pi/8$ gate) can be implemented in a 6 MBSs chain similar to that of figure 3. It relies on a parity measurement involving simultaneously both the I and Q quadratures of the cavity field (i.e. along an arbitrary angle $\phi$ in the I-Q plane), each of them being coupled to $i\hat{\gamma}_{2}\hat{\gamma}_{3}$ and $i\hat{\gamma}_{3}\hat{\gamma}_{4}$ (more details can be found in the methods section). Such a $\hat{P}_{\phi}$ measurement should be inserted in the place of the measurement of $\hat{P}_{24}$ in the sequence proposed for braiding. While such a gate is not topologically protected, it could be made exponentially accurate using mitigation techniques\cite{Oreg:19}.

\begin{figure}
    \centering
    {\small \centering\includegraphics[width=0.5\linewidth,angle=0]{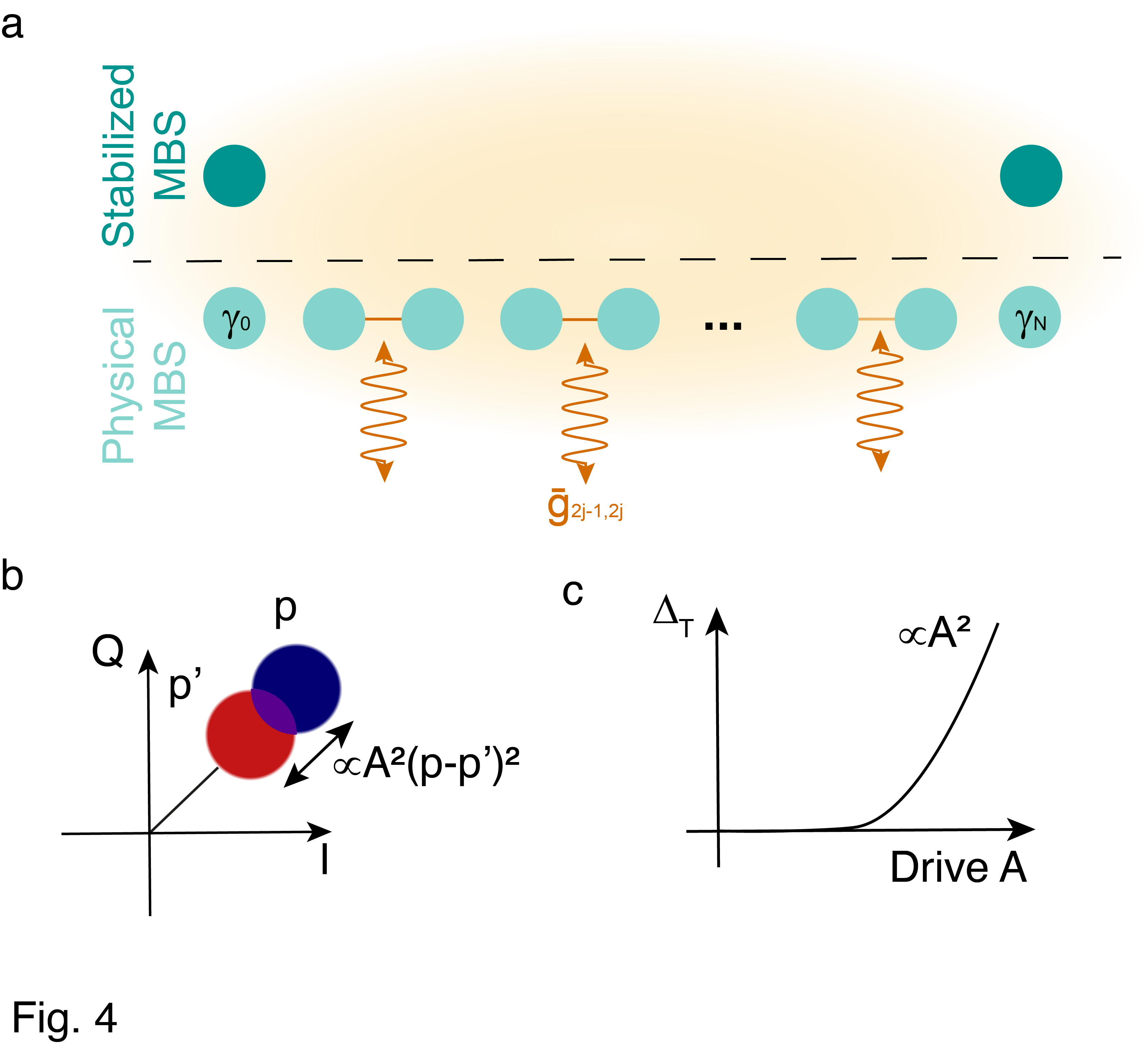}}
    \caption{\textbf{Stabilized Majorana modes : topological and polaronic gaps} \textbf{a} Schematics of the Majorana chain in cavity with gate drives inducing the stabilization. The stabilized Majorana's are sketched in dark turquoise and stem from the resulting effective Kitaev chain. \textbf{b} Sketch of the polaronic protection with  $A=\tilde{g}/2(\omega_{RF}-\omega_c)$. \textbf{c} Topological gap as a function of the drive amplitude starting from a non-topological ground state.}
    \label{fig:figure5}
\end{figure}

The previous discussion relies on the fact that we electrically manipulate, couple and read-out coupled MBSs, which seems incompatible with topological protection because of electrical noise or disorder in the $\epsilon_{jj+1}$'s. We now show another crucial consequence of the form (\ref{eq:fulleffective}) which implies that even for a chain of MBSs with finite overlap $\epsilon$ between the MBSs, one can induce with the cavity light a robust topological phase with stabilized, or self-corrected MBSs i.e with exponential protection. The principle of this exponential protection is to induce thanks to the cavity field and the gate modulation a synthetic, light-induced, Kitaev hamiltonian as sketched in figure \ref{fig:figure5} 4a. Like for error correction protocols\cite{vOppen:19}, this scheme requires some degree of redundancy and therefore longer chains than the ones considered so far. Let us first assume that we work with a chain with N MBS sections $\{0..N\}$. We assume that a gate modulation is applied every other section, starting from section $(1,2)$. In such a condition, the hamiltonian (\ref{eq:fulleffective}) becomes:
\begin{equation}
 \bar{H}_{stab}=\sum_{j odd} \frac{1}{2}\tilde{g}_{jj+1}(e^{i\phi_{jj+1}}\alpha^{*} + e^{-i\phi_{jj+1}}\alpha)i\hat{\gamma}_{j}\hat{\gamma}_{j+1}+\sum_{j} \epsilon_{jj+1}i\hat{\gamma}_{j}\hat{\gamma}_{j+1}
\end{equation}
where $\alpha$ is the static classical part of the cavity field in the rotating frame and $\epsilon_{jj+1}$ is the residual overlap between physical MBSs. Assuming that the phases $\phi_{jj+1}$ and the modulations $\tilde{g}_{jj+1}$ are tuned to $\phi$ and $\tilde{g}$ and that the phase of the coherent field $\alpha$ is $\theta$, $\bar{H}_{stab}$ can be divided into a Kitaev hamiltonian $H_K$ and a doping hamiltoninan $H_D$ and has a topological phase transition with exponentially localized MBSs at sites $0$ and $N$ (see figure 4a), for $J=\tilde{g}|\alpha|\cos{(\phi-\theta)}\gg -2\epsilon_{jj+1}$. These end MBSs are now stabilized because their overlap can be made exponentially small using macroscopic 'knobs'. The parameters $\tilde{g}$, $|\alpha|$ and $\phi-\theta$  are these 'knobs' and set the topological gap of our synthetic Kitaev hamiltonian as shown in figure 4c. This principle can be used on bigger chains to produce 4- to 6- logical Majorana chains as needed by the previously introduced protocol.

In writing the above hamiltonian, we have neglected two terms: one time dependent classical field term $\delta H^{(1)}=\tilde{g}|\alpha|\cos{(2\omega_{RF}t+\phi+\theta)}\sum_{j odd}i\hat{\gamma}_{j}\hat{\gamma}_{j+1}$ and one term arising from quantum fluctuations of the cavity field $\delta  H^{(2)}=\tilde{g}\cos{(\omega_{RF}t+\phi-\theta)}\sum_{j odd}i\hat{\gamma}_{j}\hat{\gamma}_{j+1} (\hat{b}+\hat{b}^{\dagger})$. The quantum fluctuations of the cavity field are defined by the operator $\hat{b}$. Since both perturbations are periodic in time, it is convenient to use the Floquet formalism (see Methods and Supplementary). Noting that all the parities $p_{jj+1}$ for the $(j,j+1)$ sections with $j$ odd are good quantum numbers in the Kitaev chain, the matrix elements arising in the perturbation theory depend now on $p=\sum p_{jj+1}$ which is an integer directly linked to the occupation of the chain and $m$ which is an integer arising from the Floquet ladder (see Methods and Supplementary). The first term is a fast oscillating term at roughly twice the cavity frequency. It generates matrix elements $\propto J_{\frac{m-m'}{2}}\left(\frac{(p'-p)\tilde{g}|\alpha|}{2\omega_{RF}}\right)$ (see methods). They can be safely negelected because they are of order $(\tilde{g}|\alpha|/\omega_{RF})^{\frac{|m-m'|}{2}}$ for small
$\tilde{g}|\alpha|/\omega_{RF}$  which is a very realistic condition.  It is also essential to evaluate the effect of quantum noise on the topological protection of our scheme. Defining the polaronic shift $E_0=\tilde{g}^2\omega_c/2(\omega^{2}_{RF}-\omega^{2}_c)$, we can write the quasi-energy of the driven chain as : $E_K=Jp+E_0 p^2+m \omega_c$. The result of perturbation theory on the Floquet space is twofold.  First, any local perturbation $\eta$ flipping one of the $p_{jj+1}$ can only induce an exponentially small coupling between the end stabilized Majorana's at sites $0$ and $N$ of order $\eta^{(N+1)/2}$, thus preserving the topological protection.  Second, the drive tends to shift the cavity field entangled with the state of the chain of quasienergy $E_K$ at different spots in the I-Q plane for different states of the chain with total quantum number $p$ or $p'$ because $\delta  H^{(2)}$ is a drive term proportional to $p$. The quasi-orthogonality of two coherent states with different amplitudes quenches exponentially the transition to excited states. The corresponding matrix element reads approximately: $\exp{\Big[-\frac{ \tilde{g}^2}{8\hbar^2(\omega_{RF}-\omega_c)^2}(p-p')^2\Big]}$ (equation (16) in the supplementary). This exponential polaronic protection which further protects the topological phase is presented in figure \ref{fig:figure5}4b.

In summary, we have presented circuit QED protocols based on the parametric modulation of light-matter coupling for performing advanced quantum gates for Majorana zero modes. Such an approach can also be used for a parametric stablization of the Majorana zero modes, enhancing the topological protection of a given physical platform. This should allow one to perform advanced operations with exponentially protected Majorana zero modes.

\textbf{SUPPLEMENTARY MATERIAL}

\textbf{Nearest neighbours light bonds and strong parity measurement}

We now assume that the electron-photon couplings can be locally modulated at microwave frequencies through a modulation of the Majorana quasiparticles overlap: $\epsilon_{jj+1}(t) = \bar{\epsilon}_{jj+1} + \tilde{\epsilon}_{jj+1}\cos(\omega_{RF}t + \phi_{jj+1})$ which leads to $g_{jj+1}(t) = \bar{g}_{jj+1} + \tilde{g}_{jj+1}\cos(\omega_{RF}t + \phi_{jj+1})$, $\text{ with }j$ an integer. This can be done thanks to the use of RF gates, each being capacitively coupled to one $(j,j+1)$ section of the circuit.

We specialize the discussion to section $(1,2)$. A cavity field grows in the cavity when, for example, $g_{12}$ is modulated at the cavity frequency ($\omega_{RF}=\omega_c$). It reveals directly the parity $\hat{P}_{12}$. Omitting all the other sections for the sake of simplicity, the low-energy effective Hamiltonian of the system can be reduced to:
\begin{align}
H_{el} &= i \epsilon_{12}(t)\hat{\gamma}_{1}\hat{\gamma}_{2} \label{eq:HamMajoPart}\\
H_{int} & = i g_{12}(t)\hat{\gamma}_{1}\hat{\gamma}_{2}(\hat{a}+\hat{a}^{\dagger})
\end{align}
We can rewrite the Hamiltonian in a rotating frame at $\omega_c$. We obtain:
\begin{equation}\label{eq:parity}
\tilde{H}=H_{el} + \frac{1}{2}\tilde{g}_{12}(e^{i\phi_{12}}\hat{a}^{\dagger} + e^{-i\phi_{12}}\hat{a})i\hat{\gamma}_{1}\hat{\gamma}_{2}
\end{equation}
where the static term proportional to $\bar{g}_{12}$ is neglected as a fast oscillating term, under the Rotating Wave Approximation (RWA). Equation (\ref{eq:parity}) shows an effective coupling between the Majorana's 1 and 2 which can be used to measure their parity $\hat{P}_{12}=i\hat{\gamma}_{1}\hat{\gamma}_{2}$ via the cavity field as shown in figure \ref{fig:figure2} 1. This measurement follows the same principle as the longitudinal coupling read-out for qubits\cite{Didier2015}. The parity eigenstate is read-out from the position of the coherent state spots in the I-Q plane associated to the $\hat{P}_{12}=\pm 1$ eigenvalues. The contrast for the coherent state spots in the I-Q plane is $\tilde{g}_{12}/\sqrt{\kappa}$, where $\kappa$ is the damping rate of the cavity (see Input-Output section of the methods), which can be made much larger than the width of the gaussian spots of the coherent states even deep in the topological regime where $\tilde{g}_{12} \rightarrow 0$ for small enough $\kappa$. Similarly, one can also measure $\hat{P}_{23}=i\hat{\gamma}_{2}\hat{\gamma}_{3}$. This is simply done by letting $\tilde{g}_{23}$ non zero keeping the other modulating terms negligible. This gives a concrete protocol to fuse $\hat{\gamma}_1$ and $\hat{\gamma}_2$ of figure \ref{fig:figure1}2, and to detect the fusion rules through the use of cavity photons.

\textbf{Second nearest neighbours light bonds}

We now show specifically on a 4 Majorana chain $(1,2,3,4)$ how we can obtain a second nearest neighbours photon mediated interaction between MBSs 2 and 4. Starting again from hamiltonian (\ref{eq:ham}), we now assume that the RF signal acting on the gates $(2,3)$ and $(3,4)$ is detuned from the cavity and performs the combined unitary transformation:
\begin{equation}\label{eq:RWApolar}
    U=e^{i\omega_{RF}\hat{a}^{\dagger}\hat{a}t}e^ {\Big[\frac{\tilde{g}_{23}}{\hbar(\omega_{RF}-\omega_c)}(e^{i\phi_{23}}\hat{a}^{\dagger} + e^{-i\phi_{23}}\hat{a})i\hat{\gamma}_{2}\hat{\gamma}_{3}+\frac{\tilde{g}_{34}}{\hbar(\omega_{RF}-\omega_c)}(e^{i\phi_{34}}\hat{a}^{\dagger} + e^{-i\phi_{34}}\hat{a})i\hat{\gamma}_{3}\hat{\gamma}_{4}\Big]}
\end{equation}
The first unitary transformation is the RWA in the frame of the gate drives and the second is the dispersive transformation which implies that $\omega_{RF}-\omega_c > \tilde{g}_{23},\tilde{g}_{34}$. The outcome of these two transformations is:
\begin{equation}\label{eq:gamma24}
    H=\hbar(\omega_c-\omega_{RF})\hat{a}^{\dagger}\hat{a}+8i\frac{\tilde{g}_{23}\tilde{g}_{34}}{\hbar(\omega_{RF}-\omega_c)}\sin(\phi_{23}-\phi_{34})\hat{\gamma}_{2}\hat{\gamma}_{4}(\hat{a}^{\dagger}\hat{a}+1/2)
\end{equation}

\textbf{Performing the T(or $\pi$/8)-gate}

Other useful forms of the hamiltonian (\ref{eq:fulleffective}) can be derived. A particularly important one enables the implementation of a T-gate which corresponds to a $\pi/8$ geometrical phase during the unitary evolution of the system. We specialize to the 4 MBS chain again for the sake of simplicity and assume that $\bar{\epsilon}_{34}\neq0$ and $\tilde{g}_{23}\neq 0$. For the unitary transformation $U=e^{i(\omega_{c}\hat{a}^{\dagger}\hat{a}+\bar{\epsilon}_{34}i\hat{\gamma}_{3}\hat{\gamma}_{4}t)}$ (interacting picture),  the effective hamiltonian becomes:

\begin{equation}\label{eq:Tgate1}
  H_{\pi/8}=\frac{\tilde{g}_{23}}{2}(e^{i\omega_c t}\hat{a}^{\dagger} + e^{-i \omega_c t}\hat{a})\cos(\omega_{RF}t + \phi_{23})\{\cos(2\bar{\epsilon}_{34}t)i\hat{\gamma}_{2}\hat{\gamma}_{3}+\sin(2\bar{\epsilon}_{34}t)i\hat{\gamma}_{3}\hat{\gamma}_{4}\}
\end{equation}
For $\omega_{RF}=\omega_c+2\bar{\epsilon}_{34}$, retaining only the resonant terms, we get:
\begin{equation}\label{eq:Tgate2}
  H_{\pi/8}=\frac{\tilde{g}_{23}}{2}(e^{i\phi_{23}}\hat{a}^{\dagger} + e^{-i\phi_{23}}\hat{a})i\hat{\gamma}_{2}\hat{\gamma}_{3}+\frac{\tilde{g}_{23}}{2} i(e^{i\phi_{23}}\hat{a}^{\dagger} - e^{-i\phi_{23}}\hat{a})i\hat{\gamma}_{2}\hat{\gamma}_{4}
\end{equation}
Such a form shows that the two different directions corresponding to $i\gamma_2 \gamma_3$ or ``$\sigma_x$'' and to $i\gamma_2 \gamma_4$ or``$\sigma_y$'' become coupled with the two quadratures of the cavity field (respectively I and Q). This allows us to perform a T gate simply by measuring the cavity field along the bisector between I and Q.
If such a measurement is inserted instead of the measurement of $\hat{P}_{24}$ in the braiding sequence, the unitary evolution of the wave function will pick up a $\pi/8$ geometrical phase instead of the $\pi/4$ of the braiding.

\textbf{Floquet formalism for the stabilized Majorana modes}

In deriving the effective Hamiltonian $\bar{H}_{stab}$, the cavity field has been replaced by its resonant component in the rotating frame.
As we have shown, this procedure generates a static Kitaev Hamiltonian $H_{K}$.
The goal of this section is to demonstrate that, crucially, the remarkable topological protection of Majorana edge mode degeneracy which is garanteed by $H_{K}$ also extends to the
periodically driven situation considered in the present work without relying on the rotating frame approximation.

We first write the cavity field as a sum:
\begin{equation}
\hat{a}=|\alpha|e^{-i(\omega_{RF}t+\theta)}+\hat{b}
\end{equation}
Assuming that the coupling between the Majorana chain and the cavity is modulated only on the $(j,j+1)$ bonds with $j$ odd, we get the time periodic coupling Hamiltonian:
\begin{equation}
H_{c}(t)=\sum_{j\;odd}
\left(
J+\tilde{J}\cos{(2\omega_{RF}t+\phi+\theta)}+
\tilde{g}\cos{(\omega_{RF}t+\varphi)}(\hat{b}+\hat{b}^{\dagger})
\right)i\hat{\gamma}_{j}\hat{\gamma}_{j+1}
\end{equation}
Here we have set $J=\tilde{J}\cos{\varphi}$, with
$\tilde{J}=\tilde{g} |\alpha|$ and $\varphi=\phi-\theta$.
This has the form:
\begin{equation}
H_{c}(t)=H_{K}+\delta H^{(1)}(t)+\delta H^{(2)}(t).
\end{equation}
Besides the static Kitaev Hamiltonian already derived earlier using the
rotating wave approximation, we get two time-periodic perturbations
$\delta H^{(1)}(t)$ and $\delta H^{(2)}(t)$. The former induces a time-periodic
modulation of the Kitaev coupling, $J$ being replaced by
$J_{eff}(t)=J+\tilde{J}\cos{(2\omega_{RF}t+\phi+\theta)}$. The later couples the Majorana
modes to quantum fluctuations of the cavity field.
A key feature of this model is that both $\delta H^{(1)}(t)$ and $\delta H^{(2)}(t)$
commute with $H_{K}$, and even more importantly, with its local conserved operators
$\hat{p}_{jj+1}=i\hat{\gamma}_{j}\hat{\gamma}_{j+1}$ for odd $j$.
Since the existence of
conserved local operators lies at the heart of topological protection, the persistence of this property in the full $H_{c}(t)$ is of course essential for our purpose here.

The first key ingredient to achieve topological protection is a large energy gap,
compared to the strength of the static perturbation $\epsilon_{jj+1}$. Here lies a
potential fragility of the present proposal, because inelastic interactions
due to the periodic driving may strongly reduce the value of the effective gap
below its static value $2J$.
This concern is particularly clear for the $\delta H^{(1)}(t)$ perturbation
because $J_{eff}(t)$ vanishes twice in each period $\pi/\omega_{RF}$
(or just once if $\varphi$ is an integer multiple of $\pi$).

To address this issue, we have to extend the analysis of topological
protection to situations where the reference Hamiltonian is time-periodic.
We should first understand the Floquet spectrum of $H_{c}(t)$ and then
investigate the effect of the static perturbation
$H_D=\sum_{j} \epsilon_{jj+1}i\hat{\gamma}_{j}\hat{\gamma}_{j+1}$.
To make the discussion clearer, we shall discuss separately
the Floquet spectra when either $\delta H^{(1)}(t)$ or $\delta H^{(2)}(t)$
is added to $H_{K}$.

Let us denote by $|\tau,\{p_{jj+1}\}\rangle$ a state of the Majorana chain such that:
\begin{eqnarray}
i\hat{\gamma}_{j}\hat{\gamma}_{j+1} |\tau,\{p_{jj+1}\}\rangle & = & p_{jj+1} |\tau,\{p_{jj+1}\}\rangle\;\;\;\;\;(j\;\mathrm{odd})  \\
i\hat{\gamma}_{0}\hat{\gamma}_{N} |\tau,\{p_{jj+1}\}\rangle & = & \tau |\tau,\{p_{jj+1}\}\rangle
\end{eqnarray}
Here, each eigenvalue $p_{jj+1}$ and $\tau$ can be $\pm 1$.
The Floquet eigenstates of $H_{K}+\delta H^{(1)}(t)$ have the form:
\begin{equation}
|\Psi(t)\rangle = e^{-ipJt}e^{-ip\tilde{J}\frac{\sin (2\omega_{RF}t+\phi+\theta)}{2\omega_{RF}}} |\tau,\{p_{jj+1}\}\rangle,
\label{Floquet_basis_delta_H_1}
\end{equation}
so their Floquet quasi-energy is $pJ$, which is defined modulo $\hbar \omega_{RF}$.

To study the effect of $H_D$, we view it as a perturbation of the operator
$\mathcal{L}_{1}=H_{K}+\delta H^{(1)}(t)-i\frac{d}{dt}$, acting in the Hilbert space $\mathcal{H}_{per}$ of
periodic wave-functions of $t$ with period $T$. More details on this procedure are given in the {\bf Supplementary Material} section.
From Eq.~(\ref{Floquet_basis_delta_H_1}), a complete eigenvector basis for $\mathcal{L}_{1}$ is given
by states $|\tau,\{p_{jj+1}\};m\rangle\rangle$ with eigenvalues $pJ-m\hbar\omega_{RF}$.

Topological protection means that the effective coupling between Majorana end modes generated by the static perturbation
$H_{D}$ is exponentially small in $N$. The existence of the local conserved operators
$\hat{p}_{jj+1}$ (for odd $j$) implies that such an effective coupling, proportional to $i\hat{\gamma}_{0}\hat{\gamma}_{N}$,
occurs only at order $(N+1)/2$ in perturbation theory. Indeed, the lowest order product of Majorana operators
which contains both $\hat{\gamma}_{0}$ and $\hat{\gamma}_{N}$, and which commutes with all $\hat{p}_{jj+1}$ operators (for odd $j$)
is $\prod_{l} i\hat{\gamma}_{2l}\hat{\gamma}_{2l+1}$, where $l$ runs from 0 to $(N-1)/2$. Each term of the product
corresponds to a local perturbation $i\epsilon_{2l,2l+1}\hat{\gamma}_{2l}\hat{\gamma}_{2l+1}$. Let us assume that
it connects state $|\tau,\{p_{jj+1}\};m\rangle\rangle$ to state $|\tau',\{p'_{jj+1}\};m'\rangle\rangle$.
In this case, one has $p'_{jj+1}=\pm p_{jj+1}$, the minus sign occurring only if $j=2l-1$ or $j=2l+1$. To each intermediate state
is associated an energy denominator $(p_{GS}-p)J+m\hbar\omega_{RF}$, where $p_{GS}=-(N-1)/2$ since
$p_{j,j+1}=-1$ for any odd $j$ in any of the two-fold degenerate ground-states of $H_{K}$. Compared to the static case,
we see that the large gap proportional to $J$ is replaced by the smaller value $\min_{\{m\}} (J-m\hbar\omega_{RF})$. Therefore, a necessary condition for topological protection to survive in the presence of a periodic modulation of $J_{eff}$ is that
inelastic transitions to states with a non-zero value of $m$ should be strongly suppressed. It is thus crucial to examine in more detail the matrix elements of the perturbation.

Using the time dependence of unperturbed Floquet eigenstates given by Eq.~(\ref{Floquet_basis_delta_H_1}), we get:
\begin{equation}
\langle \langle \tau',\{p'_{jj+1}\};m' |i \hat{\gamma}_{2l}\hat{\gamma}_{2l+1} |\tau,\{p_{jj+1}\};m \rangle \rangle \propto
J_{\frac{m-m'}{2}}\left(\frac{(p'-p)\tilde{J}}{2\omega_{RF}}\right), \mathrm{for}
\;\;\; m'-m \;\mathrm{even},
\label{matrix_element_1}
\end{equation}
and this matrix element vanishes if $m'-m$ is odd. In Eq.~(\ref{matrix_element_1}),
$J_{\frac{m-m'}{2}}$ is the usual Bessel function of the first kind.
Since the above matrix element is proportional to
$(\tilde{J}/\omega_{RF})^{\frac{|m-m'|}{2}}$ at small
$\tilde{J}/\omega_{RF}$, we see that inelastic transitions to states with a non-zero value of $m$ are suppressed when $\tilde{J} << \omega_{RF}$, i.e. when the driving frequency is large compared to the time averaged gap of the effective Kitaev chain.

Although this argument is quite compelling,
a potential danger lies in the fact that the ordering between the $(N+1)/2$
local perturbations $i\epsilon_{2l,2l+1}\hat{\gamma}_{2l}\hat{\gamma}_{2l+1}$ is arbitrary, so we have $((N^2-1)/8)!$ terms at order $(N+1)/2$. In the static case,
this factorial growth is compensated by the large value of typical energy denominators.
In the periodically modulated case, no exact solution in the presence of the static
perturbation $H_D$ is available, and to establish rigorously that the effective coupling between boundary Majorana modes decays exponentially with $N$ would require
a more involved analysis, which is beyond the scope of the present work.

Let us now turn to the Floquet eigenstates of $H_{K}+\delta H^{(2)}(t)$.
Since this Hamiltonian commutes with the conserved operators of $H_K$,
we can put the Majorana chain in one of the states $|\tau,\{p_{jj+1}\}\rangle$
for all times $t$. The quantum oscillator mode of the cavity if then subjected to the
Hamiltonian:
\begin{equation}
H_{cav}(t)=\hbar\omega_{c}\hat{b}^{\dagger}\hat{b}+p\tilde{g}\cos(\omega_{RF}t+\phi)\:
(\hat{b}+\hat{b}^{\dagger})
\label{def_H_cav_(t)}
\end{equation}
The Floquet spectrum of $H_{cav}$ is discussed in the {\bf Supplementary Material} section.
Let us first consider the non-resonant case, when the detuning $\delta=
\omega_c-\omega_{RF}$ is larger than the cavity damping rate $\Gamma$.
Combining the Majorana chain and the cavity, the eigenstates of the operator
$\mathcal{L}_{2}=H_{K}+\delta H^{(2)}(t)-i\frac{d}{dt}$, acting in the Hilbert space $\mathcal{H}_{per}$ can be written as $|\tau,\{p_{jj+1}\};n;m\rangle\rangle$, where
$n$ is a non-negative integer associated to the cavity oscillator and, as before,
$m$ labels Fourier modes in the auxiliary space of periodic functions of time.
The corresponding eigenvalues are $pJ+p^{2}E_{0}+n\hbar\omega_{c}-m\hbar\omega_{RF}$.

The effect of the static perturbation $H_D$ is similar to the previous case. The local
perturbation term $i\epsilon_{2l,2l+1}\hat{\gamma}_{2l}\hat{\gamma}_{2l+1}$ acts only on the Majorana chain, where
it connects state $|\tau,\{p_{jj+1}\};m\rangle\rangle$ to state $|\tau',\{p'_{jj+1}\};m'\rangle\rangle$. The new feature with $\delta H^{(2)}(t)$,
compared to $\delta H^{(1)}(t)$, is that the transition between these two states of the
chain also modifies the amplitude of the periodic driving seen by the cavity mode.
The analysis of these matrix elements is presented in the \textbf{Supplementary Material} section
in the case where the detuning $\delta$ is small. One of the main features is the approximate selection rule $n'-m'=n-m$. This implies that the energy denominators
$(p_{GS}-p)J+(p_{GS}^{2}-p^{2})E_{0}-n\hbar\omega_{c}+m\hbar\omega_{RF}$
in the leading contributions to the effective coupling between boundary Majorana modes
are close to the values $(p_{GS}-p)J$ governing the static case. This gives strong support to our claim that topological protection is achieved in this model of a driven
Majorana chain. Another bonus provided by the driven model comes from the
fact that the matrix elements of $H_D$ are proportional to the overlap between coherent
states:
\begin{equation}
 \langle z'(t)|z(t) \rangle=\exp\left(-\frac{(p'-p)^{2}\tilde{g}^{2}}{8\:\hbar^2\delta^{2}}
 \right).
\end{equation}
The gaussian factor in Eq. (19) may be significantly smaller than 1, which would enhance the protection of the ground-state degeneracy with respect to residual static perturbations
such as $H_D$. This is analogous to the reduction of a polaron hopping amplitude, due to
its strong coupling to lattice vibration modes. In this analogy, the polaron becomes
the Majorana chain and the vibration modes are replaced by the cavity oscillator.

In the case of a finite cavity damping $\kappa$, it is necessary to take into account
the coupling of the cavity oscillator to a continuum of environmental modes.
In the limit of a small damping $\omega_{c} \gg \kappa$, it is shown in
the \textbf{Supplementary Material} section that the driving term in $H_{cav}(t)$
couples mostly to the dressed modes near the cavity frequency $\omega_{c}$.
Therefore, most of the previous analysis of topological protection in the limit of
small detuning survives in the case of a small but finite damping.

\textbf{Input-Output theory}

We show here how one can capture the cavity based measurement processes for the parity of the Majorana chain using an input-output theory. This method gives results which agree with equation (2) of the main text but it also allows to capture the dissipative dynamics related to the projective measurement of the system. The equations of motion for the photonic field in the cavity $\hat{a}$ with loss rate $\kappa$ and for the input and output fields, $\hat{a}_{in}$ and $\hat{a}_{out}$, read, for a pair of MBSs $(1,2)$:
\begin{align}
&\frac{d\hat{a}}{dt} = \frac{-i}{\hbar}[\hat{a},H]-\frac{\kappa}{2}\hat{a}\\
&\quad= \frac{-i\tilde{g}_L}{2} \hat{P}_{12} - \frac{\kappa}{2} \hat{a} -\sqrt{\kappa}\hat{a}_{in}\\
&\hat{a}_{out}=\hat{a}_{in}+\sqrt{\kappa}\hat{a}
\end{align}
with $\hat{P}_{12}(t)$ constant since $[H, \hat{P}_{12}]=0$.

In the semi-classical regime ($\alpha = <\hat{a}>$) and in absence of a cavity drive, the output field is given by:
\begin{equation}\label{eq:alphaout}
\alpha_{out}(t)=\frac{-i\tilde{g}_{12}}{\sqrt{\kappa}}<\hat{P}_{12}>(1-e^{-\kappa t/2}).
\end{equation}

The modulation of $g_{12}$ populates the cavity field, as represented in figure 1b. By measuring the occupation of the cavity along the appropriate quadrature (more specifically, with a measurement phase of $\phi_{meas} = \phi_{12} + \pi/2\,[\pi]$), one can therefore perform a measurement of the parity. The SNR in this measurement depends on $\frac{\tilde{g}_{12}}{\sqrt{\kappa}}$ and on the measurement time $\tau$. It is given by \cite{Didier2015}:
\begin{equation}
\mathrm{SNR} = \frac{\sqrt{8}\lvert \tilde{g}_{12}\rvert}{\kappa}\sqrt{\kappa\tau}(1-\frac{2}{\kappa\tau}(1-e^{-\kappa\tau/2}))
\end{equation}

We now  show how the parity $\hat{P}_{24}$ can be measured as a dispersive shift of the cavity resonant frequency as explained in the main text, using:
\begin{equation}
    g_{23(34)}(t)=\bar{g}_{23(34)} + \tilde{g}_{23(34)}\cos(\omega_{RF}t+\phi_{23(34)})
\end{equation}.

The coupled equations of motion are:

\begin{align}
    &\frac{d\hat{a}}{dt} = -i\omega_c\hat{a}-ig_{23}(t)\hat{\gamma}_2\hat{\gamma}_3 - ig_{34}(t)\hat{\gamma}_3\hat{\gamma}_4 - \kappa/2\\
    &\frac{d\hat{\gamma}_2\hat{\gamma}_3}{dt} = -2i g_{34}(t)\hat{\gamma}_2\hat{\gamma}_4(\hat{a}+\hat{a}^\dagger) \\
    &\frac{d\hat{\gamma}_3\hat{\gamma}_4}{dt} = 2i g_{23}(t)\hat{\gamma}_2\hat{\gamma}_4(\hat{a}+\hat{a}^\dagger)
\end{align}

In the rotating frame (at $\omega_c$), and keeping resonant terms in the RWA, we get a first reduced equation on the cavity field:
\begin{equation}
\frac{d\hat{a}}{dt} = -\frac{\kappa}{2}\hat{a} -2\int d\tau \hat{\gamma}_2(\tau)\hat{\gamma}_4(\tau)(g_{23}(t)g_{34}(\tau)-g_{34}(t)g_{23}(\tau))(\hat{a}+\hat{a}^\dagger)
\end{equation}

We additionally suppose $\lvert \omega_{RF} - \omega_c \rvert << \omega_c$ so that we can neglect the time evolution of $\hat{\gamma}_2\hat{\gamma}_4$, as well as the one of $\hat{a}$ in the above integral. This gives:

\begin{align}
   &\frac{d\hat{a}}{dt}= - 4\hat{\gamma}_2\hat{\gamma}_4\hat{a}\frac{\tilde{g}_{23}\tilde{g}_{34}}{\omega_c - \omega_{RF}}\sin(\phi_{34} - \phi_{23}) - \frac{\kappa}{2}\hat{a}
\end{align}
One sees here again that the optimum is $\phi_{34} - \phi_{23}=\pi/2$.

\textbf{Experimental requirements and distinction with accidental Andreev bound states}
Let us now estimate the feasibility of this scheme, and more specifically whether it could indeed allow for single-shot readout of the parity.
With coplanar waveguide (CPW) resonators, cavity loss can be extremely low. However, in the scenario considered we also need to measure the cavity output in a time much smaller compared to the parity lifetime. Since the measurement time is of the order of a few $\kappa$, we assume a loss of $\kappa = 2\pi \times 1 MHz$ to be conservative (we thus require more than tens of microseconds for the parity lifetime). We then need to estimate how strongly the coupling strength can be modulated. In electrical circuits, coupling strength between a charge and a cavity of the order of $g = 2\pi\times 100 MHz$ are now achievable \cite{Stockklauser2017}, and would still be compatible with the condition $ \epsilon_{jj+1}, g_{jj+1} \ll \omega_c $. Assuming that this coupling strength can by modulated by a factor $10\%$, either by modulating the position of the Majorana pair or by modifying the shape of the electromagnetic mode, our scheme enables single-shot cavity readout of the parity.

It is important to stress that our setup can also distinguish between Majorana modes and accidental Andreev bound states. Whereas braiding is a priori the most unambiguous way of distinguishing between accidental Andreev bound states and Majorana modes, establishing the fusion rules should be enough for a large class of situations. An Andreev bound state is expected to give rise to a transverse coupling which yields a trajectory of the type of figure 1e. Measuring several sections of the chain which display only trajectories of the type of figure 1c in the I-Q plane (i.e. longitudinal coupling) in the fusion rule setup (with 4 nodes) should constraint very much the models with accidental Andreev bound states (if any exists) yielding the same signature.

\textbf{State sequence in the braiding protocol}\label{sec:crosstalk}

We recall first the measurement based braiding protocol and specifically apply it to our scheme.
Let us consider again a linear chain of four Majorana quasiparticles, $\hat{\gamma}_{j=1..4}$. The set of measurements needed for performing a braiding operation $\hat{B}_{14}$ between two Majorana $\hat{\gamma}_{1}$ and $\hat{\gamma}_{4}$ stems from the identity \cite{Bonderson2008,Vijay2016}:
\begin{align}
& \hat{\Pi}_{23}\hat{\Pi}_{21}\hat{\Pi}_{24}\hat{\Pi}_{23} = \frac{1}{\sqrt{8}} \hat{\Pi}_{23} \hat{B}_{14} \label{Eq:Braiding} \\
& \hat{\Pi}_{jk} = \frac{1}{2}(1+\hat{P}_{jk})
\end{align}
The operator $\hat{\Pi}_{jk}$ projects the electronic state onto the subspace with parity $\hat{P}_{jk}=1$.

For the state sequence presented in this work, we start by measuring $\hat{P}_{46}$ giving an intial state
$\ket{\Psi_{init}} = \frac{1}{\sqrt{2}}(\ket{1_{12}1_{34}0_{56}}+i\ket{1_{12}0_{34}1_{56}})$, postselecting the $\ket{0_{46}}$ parity state, starting from $\ket{1_{12}1_{34}0_{56}}$.

Then, for the state sequence of $\hat{B}_{41}$, we project onto the $+1$ eigenvalue using the projection operators $\hat{\Pi}_{23}$, $\hat{\Pi}_{21},\hat{\Pi}_{24}$ and then $\hat{\Pi}_{23}$. This gives the sequence of states:
\begin{align}\nonumber
\ket{\Psi_{I}}=\hat{\Pi}_{23}\ket{\Psi_{init}}= \frac{1}{2}(\ket{0_{12}0_{34}0_{56}}+\ket{1_{12}1_{34}0_{56}}+i\ket{1_{12}0_{34}1_{56}}+i\ket{0_{12}1_{34}1_{56}} )\\\nonumber
\ket{\Psi_{II}}=\hat{\Pi}_{21}\ket{\Psi_{I}}= \frac{1}{\sqrt{2}}( \ket{1_{12}1_{34}0_{56}}+i\ket{1_{12}0_{34}1_{56}})\\\nonumber
\ket{\Psi_{III}}=\hat{\Pi}_{24}\ket{\Psi_{II}}= \frac{1}{2}(i\ket{0_{12}0_{34}0_{56}}+\ket{1_{12}1_{34}0_{56}}+i\ket{1_{12}0_{34}1_{56}}+\ket{0_{12}1_{34}1_{56}})\\\nonumber
\ket{\Psi}_{\text{braided 41}}=\ket{\Psi_{IV}}=\hat{\Pi}_{23}\ket{\Psi_{III}}= \frac{e^{i\pi/4}}{2}(\ket{0_{12}0_{34}0_{56}}+\ket{1_{12}1_{34}0_{56}}+\ket{1_{12}0_{34}1_{56}} +\ket{0_{12}1_{34}1_{56}})\\\nonumber
\end{align}

The state sequence for $\hat{B}_{14}$ is:
\begin{align}\nonumber
\ket{\Psi_{I}}=\hat{\Pi}_{23}\ket{\Psi_{init}}= \frac{1}{2}(\ket{0_{12}0_{34}0_{56}}+\ket{1_{12}1_{34}0_{56}}+i\ket{1_{12}0_{34}1_{56}}+i\ket{0_{12}1_{34}1_{56}} )\\\nonumber
\ket{\Psi_{II}}=\hat{\Pi}_{24}\ket{\Psi_{I}}=\frac{e^{i\pi/4}}{2}(\ket{0_{12}0_{34}0_{56}}-i\ket{1_{12}1_{34}0_{56}}+i\ket{1_{12}0_{34}1_{56}} +\ket{0_{12}1_{34}1_{56}})\\\nonumber
\ket{\Psi_{III}}=\hat{\Pi}_{12}\ket{\Psi_{II}}= \frac{e^{-i\pi/4}}{2}(\ket{1_{12}1_{34}0_{56}}-\ket{1_{12}0_{34}1_{56}})\\\nonumber
\ket{\Psi}_{\text{braided 14}}=\ket{\Psi_{IV}}=\hat{\Pi}_{23}\ket{\Psi_{III}}= \frac{e^{-i\pi/4}}{2}(\ket{0_{12}0_{34}0_{56}}+\ket{1_{12}1_{34}0_{56}}-\ket{1_{12}0_{34}1_{56}} -\ket{0_{12}1_{34}1_{56}})\\\nonumber
\end{align}
We therefore arrive at the result of the main text : $\ket{\Psi}_{\text{braided 14}}$ is an eigenvector of $\hat{P}_{45}$ for the clockwise braiding with eigenvalue $-1$, yielding the blue spot in the I-Q plane and $\ket{\Psi}_{\text{braided 41}}$ is an eigenvector of $\hat{P}_{45}$ for the anti-clockwise braiding with eigenvalue $+1$, yielding the red spot in the I-Q plane. The reasoning for the even total parity is exactly the same.

 \textbf{Perturbation theory for Floquet Hamiltonians}

 Let us consider a time-periodic Hamiltonian $H(t)=H_{0}(t)+\epsilon H_{1}(t)$.
 Its time period is $T=2\pi/\omega_{RF}$, where $\omega_{RF}$ is the driving frequency.
 We start with a Floquet eigenstate $ \ket{\Psi (t)}=e^{-iE_{0}t/ \hbar}\ket{\chi_{0}(t)}$ for the
 unperturbed Hamiltonian $H_{0}(t)$. The Floquet energy $E_{0}$ is defined modulo
 $\hbar \omega_{0}$, and the state $\ket{\chi(t)}$ is periodic in $t$ with period $T$.
 Perturbation theory in $\epsilon$ is more conveniently implemented within the infinite dimensional Hilbert space
 $\mathcal{H}_{per}$
 of time-periodic wave-functions$\ket{\chi(t)}$. $\mathcal{H}_{per}$ can be described as the tensor product of
 the physical Hilbert space $\mathcal{H}_{phys}$ and the space of $T$-periodic functions of $t$.
 We first notice that $|\chi_{0}(t)\rangle$ is an eigenvector
 of the operator $\mathcal{L}_{0}=H_{0}(t)-i\frac{d}{dt}$ acting in $\mathcal{H}_{per}$,
 with the eigenvalue $E_{0}$. Determining Floquet eigenstates for the full Hamiltonian $H(t)$ is equivalent
 to finding eigenvectors of $\mathcal{L}=\mathcal{L}_{0}+ \epsilon H_{1}(t)$ in $\mathcal{H}_{per}$.
 Since $\mathcal{L}$ is a hermitian operator for the hermitian scalar product in $\mathcal{H}_{per}$ defined by
 $\langle\langle \chi_{1}|\chi_{2}\rangle\rangle=\frac{1}{T}\int_{0}^{T} \langle \chi_{1}(t)|\chi_{2}(t)\rangle$,
 we can use the standard procedure of time-independent perturbation theory for $\mathcal{L}$. Doing this, we have to keep in
 mind that each Floquet eigenstate $|\chi(t)\rangle$ of $\mathcal{L}$ with eigenvalue $E$
 generates an infinite ladder of Floquet eigenstates $|\chi;m\rangle\rangle$ of $\mathcal{L}$ with eigenvalues
 $E - m \hbar\omega_{RF}$. The notation emphasizes that $|\chi;m\rangle\rangle$ should be regarded as a vector in
 $\mathcal{H}_{per}$, and the corresponding time-periodic wave-function is
 $|\chi_{m}(t)\rangle= e^{-im\omega_{RF}t}|\chi(t)\rangle$.

 \textbf{Floquet spectrum for a periodically driven oscillator}

We consider a single oscillator mode with a time-periodic and linear driving,
described by the Hamiltonian~(\ref{def_H_cav_(t)}). To simplify notations, we shall
assume that $\phi=0$. The dynamics of this driven
oscillator is easily solved by considering the dressed annihilation operator
$\hat{b}(t)$, solution of the evolution equation
$i\frac{d}{dt}\hat{b}(t)=[H(t),\hat{b}(t)]$, and subjected to the quasi-periodicity
condition $\hat{b}(t+T)=e^{i\omega_{c}T}\,\hat{b}(t)$.
Note that this time-dependent operator should not be confused with the Heisenberg
picture of the $\hat{b}$ operator. A simple calculation shows that
$\hat{b}(t)=e^{i\omega_{c}t}\,(\hat{b}+z(t))$, with:
\begin{equation}
z(t)=\frac{p\tilde{g}}{2\hbar}\left(\frac{e^{-i\omega_{RF}t}}{\omega_{c}-\omega_{RF}}
+ \frac{e^{i\omega_{RF}t}}{\omega_{c}+\omega_{RF}}\right).
\label{eq_def_z(t)}
\end{equation}
We introduce the time-periodic stationary state $|\mathcal{S}(t)\rangle$, which
is a solution of the Schr\"odinger equation, such that it is annihilated by
$\hat{b}(t)$ at all time $t$. $|\mathcal{S}(t)\rangle$ is thus a coherent state, and more
precisely:
\begin{equation}
 |\mathcal{S}(t)\rangle=e^{-i\frac{p^{2}E_{0}}{\hbar}\left(t+\frac{\sin (2\omega_{RF}t)}{2\omega_{RF}}\right)}e^{-\frac{|z(t)|^{2}}{2}}
 e^{-z(t)\hat{b}^{\dagger}}|0\rangle=e^{-ip^{2}\beta(t)}|z(t)\rangle,
 \label{expression_stationary_state}
\end{equation}
with $E_{0}=\frac{\tilde{g}^{2}\omega_{c}}{2\hbar(\omega_{RF}^{2}-\omega_{c}^{2})}$.
The complete Floquet spectrum of this driven oscillator is given by the states
$(\hat{b}^{\dagger}(t))^{n}|\mathcal{S}(t)\rangle$, $n$ non-negative integer,
whose Floquet quasi-energies are $E_{n}=p^{2}E_{0}+n\hbar\omega_{c}$.

\textbf{Matrix elements of $H_{D}$ between eigenstates of $\mathcal{L}_{2}$}

We need to evaluate $\langle \langle \tau',\{p'_{jj+1}\};n';m' |i \hat{\gamma}_{2l}\hat{\gamma}_{2l+1} |\tau,\{p_{jj+1}\};n;m \rangle \rangle$.
This is proportional to:
\begin{equation}
I=\frac{1}{T_{D}}\int_{0}^{T_{D}}e^{i\left((m'-m)\omega_{RF}t+(p'^{2}-p^{2})
\beta(t)\right)}\langle z'(t)|(\hat{b}+z'(t))^{n'}(\hat{b}^{\dagger}+\bar{z}(t))^{n}|z(t) \rangle
\label{def_I_integral}
\end{equation}
Here $z(t)$ is defined in Eq.~(\ref{eq_def_z(t)}), and $z'(t)$ is obtained after replacing $p$ by $p'$. The matrix element in the integral can be expressed as
the overlap $\langle z'(t)|z(t) \rangle$ between two coherent states times a polynomial
\begin{equation}
P_{n',n}(z'-z,\bar{z}-\bar{z}')=\sum_{l=0}^{\min(n',n)}\frac{n'!\;n!}{l!\,(n'-l)!\,(n-l)!}(z'-z)^{n'-l}\,(\bar{z}-\bar{z}')^{n-l}.
\end{equation}
Since the ratio $z'(t)/z(t)=p'/p$ is a real number, the overlap
$\langle z'(t)|z(t) \rangle$ is also real. Let us concentrate on the time evolution
of phase factors entering the integral~(\ref{def_I_integral}). For arbitrary values
of the detuning $\delta$, it is difficult to make precise statements. So let us assume
now that $\delta$ is still finite, but small. Therefore, the first term in the right-hand side of Eq.~(\ref{eq_def_z(t)}) dominates, and the argument of $z(t)$ is close to $-\omega_{RF}t$. If the driving amplitude $\tilde{g}/\hbar$ is sufficiently small, so that
the quadratic correction $\beta(t)$ (whose time-averaged value is zero) can also be neglected, we see that the argument
in the R.H.S. of~(\ref{def_I_integral})  is well approximated by $(m'-m-n'+n)\omega_{RF}t$. So we see that we have an approximate selection rule for the
matrix elements of the static perturbation between eigenstates of the
$\mathcal{L}_{2}$ operator:
\begin{equation}
n'-m'=n-m
\label{approximate_selection_rule}
\end{equation}

\textbf{Floquet spectrum in the presence of cavity damping}

At resonance ($\omega_{c}=\omega_{RF}$), the stationary state $|\mathcal{S}(t)\rangle$
is ill-defined, because $z(t)$ defined in Eq.~(\ref{eq_def_z(t)}) becomes infinite.
To recover a stationary state, we need to take into account cavity damping.
A simple way to model such a situation is to consider the following Hamiltonian:
\begin{equation}
    H=\omega_{0}\,\hat{b}^{\dagger}\hat{b}+\int_{0}^{\infty}d\omega\,\left(
    \omega \,\hat{c}^{\dagger}(\omega)\hat{c}(\omega) +
    x(\omega) \,\hat{b}^{\dagger}\hat{c}(\omega)+
    \bar{x}(\omega) \,\hat{c}^{\dagger}(\omega)\hat{b}
    \right),
\end{equation}
where the creation and annihilation operators $\hat{c}^{\dagger}(\omega)$,
$\hat{c}(\omega)$ refer to the continuum of environmental modes, and $x(\omega)$ is an arbitrary coupling function. This Hamiltonian can be diagonalized using
dressed operators $\hat{\gamma}^{\dagger}(\omega)$,
$\hat{\gamma}(\omega)$, which satisfy
$[H,\hat{\gamma}^{\dagger}(\omega)]=\omega\,\hat{\gamma}^{\dagger}(\omega)$.
The cavity oscillator operator $\hat{b}$ can be written as a linear combination
of the dressed operators $\hat{\gamma}(\omega)$:
\begin{equation}
\hat{b}= \int_{0}^{\infty}d\omega\,\frac{x(\omega)}{\omega-\omega_{0}-
\Sigma_{R}(\omega)}\,\hat{\gamma}(\omega),
\label{spectral_dec}
\end{equation}
where $\Sigma_{R}(\omega)$ is the retarded self-energy defined by:
\begin{equation}
\Sigma_{R}(\omega)= \int_{0}^{\infty}d\omega'\,\frac{|x(\omega')|^{2}}
{\omega-\omega'+i\eta},\;\;\;\;\eta \rightarrow 0^{+}.
\end{equation}

When the cavity lifetime $\tau=\kappa^{-1}$ is long ($\omega_{c}\tau  \gg 1$), the spectral
decomposition~(\ref{spectral_dec}) is peaked around the cavity frequency $\omega_{c}$.
In the vicinity of $\omega_{c}$, we can Taylor expand the self energy, which
gives:
\begin{equation}
\Sigma_{R}(\omega) \simeq \omega_{c}-\omega_{0}+(1-Z^{-1})(\omega-\omega_{c}+i\kappa/2).
\end{equation}
Denoting by $u(\omega)$ the coefficient of $\hat{\gamma}(\omega)$ in the spectral
decomposition~(\ref{spectral_dec}), we see that
$u(\omega)\simeq Zx(\omega_{c})/(\omega-\omega_{c}+i\kappa/2)$.

In the presence of the periodic driving described by the Hamiltonian~(\ref{def_H_cav_(t)}), each dressed mode of the continuum reaches its
own stationary state $|\mathcal{S}\rangle_{\omega}$, constructed by replacing
$\hat{b}^{\dagger}$ by $\hat{\gamma}^{\dagger}(\omega)$ and multiplying
$\tilde{g}/\hbar$ by the factor $u(\omega)$ in Eq.~(\ref{expression_stationary_state}).
Since $u(\omega)$ is peaked around the cavity frequency $\omega_{c}$ when
$\omega_{c}\tau  \gg 1$, we see that the driving term couples mostly to the dressed
modes near the cavity frequency $\omega_{c}$.

{\bf Acknowledgements} We would like to thank J. Klinovaja,  F. von Oppen and N. Regnault for fruitful discussions. This work was supported by the Quantera project SuperTop. Correspondence and requests for materials should be addressed to A.C. (audrey.cottet@ens.fr) or T.K. (takis.kontos@ens.fr).

\end{document}